\documentclass[prl,a4paper,showpacs,twocolumn,twoside]{revtex4}
\usepackage{amsmath,graphicx}
\usepackage[T1]{fontenc}

\begin{document}
\bibliographystyle{apsrev}
\title{Unphysical Predictions of Some Doubly Special Relativity Theories}
\author{Jakub Rembieli{\'n}ski}
\email{J.Rembielinski@merlin.fic.uni.lodz.pl}
\affiliation{Katedra Fizyki Teoretycznej, Uniwersytet {\L}{\'o}dzki\\
  ul.~Pomorska 149/153, 90-236 {\L}{\'o}d{\'z}, Poland}
\author{Kordian A. Smoli{\'n}ski}
\email{K.A.Smolinski@merlin.fic.uni.lodz.pl}
\affiliation{Katedra Fizyki Teoretycznej, Uniwersytet {\L}{\'o}dzki\\
  ul.~Pomorska 149/153, 90-236 {\L}{\'o}d{\'z}, Poland}
\begin{abstract}
  A kind of doubly special relativity theory proposed by J. Magueijo
  and L. Smolin [Phys.\ Rev.\ Lett.\ \textbf{88}, 190403 (2002)] is
  analysed.  It is shown that this theory leads to serious physical
  difficulties in interpretation of kinematical quantities.  Moreover,
  it is argued that statistical mechanics and thermodynamics cannot be
  resonably formulate within the model proposed in the mentioned
  paper.
\end{abstract}
\pacs{03.30.+p, 05.20.-y, 05.70.-a}
\maketitle

Recently J.~Magueijo and L. Smolin \cite{magueijo02} have proposed a
modification of special relativity with intrinsically built in Planck
length.  Their construction is based on a non-linear form of the
Lorentz group action in the momentum space.  This is related to a
special choice of the Lie algebra of the Lorentz group in the
enveloping algebra of the nilpotent Lie algebra generated by $\partial/\partial
p_\mu$, $p_\mu$ and $I$.  Accordingly to \cite{magueijo02} the Lorentz
boosts in the $x$-direction read
\begin{subequations}
  \label{eq:1}
  \begin{align}
    \label{eq:1a}
    p_0' &= \frac{p_0 \cosh \xi + p_1 \sinh \xi}{1 + l p_0 (\cosh \xi - 1)
      + l p_1 \sinh \xi},\\
    \label{eq:1b}
    p_1' &= \frac{p_1 \cosh \xi + p_0 \sinh \xi}{1 + l p_0 (\cosh \xi - 1)
      + l p_1 \sinh \xi},\\
    \label{eq:1c}
    p_2' &= \frac{p_2}{1 + l p_0 (\cosh \xi - 1) + l p_1 \sinh \xi},\\
    \label{eq:1d}
    p_3' &= \frac{p_3}{1 + l p_0 (\cosh \xi - 1) + l p_1 \sinh \xi},
  \end{align}
\end{subequations}
where $l$ is interpreted as the Planck length $l_P$.  The above
transformations together with rotations close to the Lorentz group and
leave unchanged the following invariant built from the particle
four-momentum $p_\mu$:
\begin{equation}
  \label{eq:2}
  m^2 c^2 = \frac{\eta^{\mu\nu} p_\mu p_\nu}{(1 - l p_0)^2},
\end{equation}
when $c$, as usually, denotes the speed of light.

The scheme proposed in \cite{magueijo02} belongs to the class of
recently investigated models known as ``doubly special relativity''
\cite{magueijo02,amelino01a,amelino01b,amelino01c,amelino02,lukierski02,%
kowalski02a,kowalski02b}.  %
In this paper we analyze this scheme in details and we show that it is
plagued by physically unacceptable features.

To present our discussion as simply as possible, in the following we
restrict ourselves to the two-dimensional case.  We will denote the
energy and momentum as $c p_0$ and $p$, respectively.  Furthermore, to
separate the dimensional and scale parameters of the model we use the
quantity $\lambda/m_P c$ instead of $l$, where $\lambda$ is dimensionless and $m_P
= \sqrt{\hbar c/G_N}$ is the Planck mass.  It is enough to consider
positive values of $\lambda$.  Therefore the Lorentz transformations
\eqref{eq:1} take the form
\begin{subequations}
  \label{eq:3}
  \begin{align}
    p_0' &= \frac{p_0 \cosh \xi + p \sinh \xi}{1 + \frac{\lambda}{m_P c} p_0
      (\cosh \xi - 1) + \frac{\lambda}{m_P c}  p\sinh \xi},\\
    p' &= \frac{p_0 \sinh \xi + p \cosh \xi}{1 + \frac{\lambda}{m_P c} p_0
      (\cosh \xi - 1) + \frac{\lambda}{m_P c} p \sinh \xi},
  \end{align}
\end{subequations}
while the invariant~\eqref{eq:2} reads
\begin{equation}
  \label{eq:4}
  \frac{p_0^2 - p^2}{\left(1 - \frac{\lambda}{m_P c} p_0\right)^2} = m^2 c^2.
\end{equation}

Let us consider firstly the possible solutions of~\eqref{eq:4}.  It
follows that we have singularity at $p_0 = m_P c/\lambda$, so possible range
of $p_0$ is restricted to the two distinct areas: $p_0 < m_P c/\lambda$ or
$p_0 > m_P c/\lambda$.

Now, for $p_0 \neq m_P c/\lambda$ Eq.~\eqref{eq:4} determines algebraic conics;
namely
\begin{subequations}
  \label{eq:5}
  \begin{align}
    &\text{hyperbola}, &\text{when } m &< m_P/\lambda,\\
    &\text{parabola}, &\text{when }  m &= m_P/\lambda,\\
    &\text{ellipse}, &\text{when } m &> m_P/\lambda,\\
    &\text{the pair of half-lines}, &\text{when } m &= 0.
  \end{align}
\end{subequations}

\begin{figure*}
  \centering
  \begin{tabular}{cccc}
    (a)&(b)&(c)&(d)\\
    \includegraphics[width=.24\textwidth]{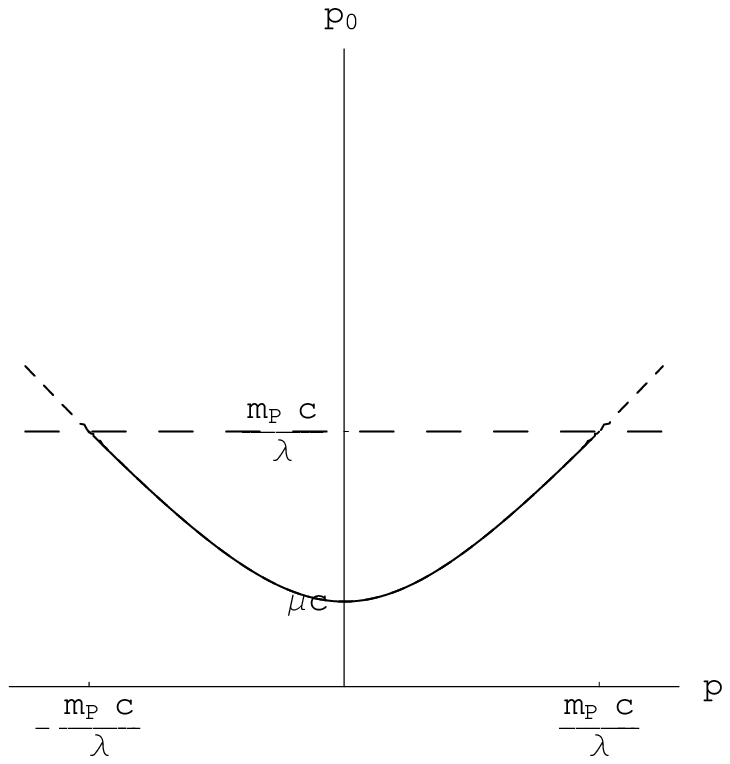}&
    \includegraphics[width=.24\textwidth]{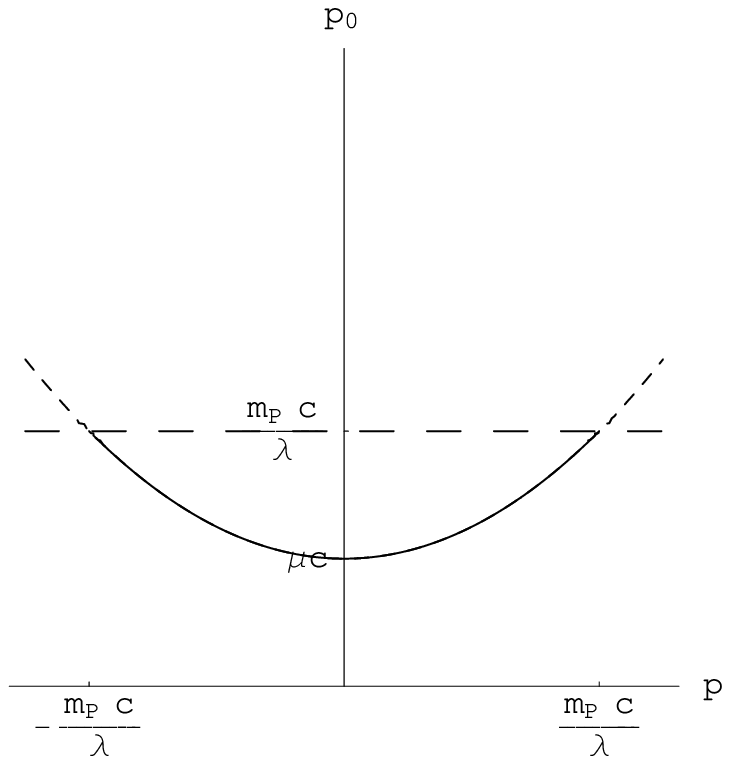}&
    \includegraphics[width=.24\textwidth]{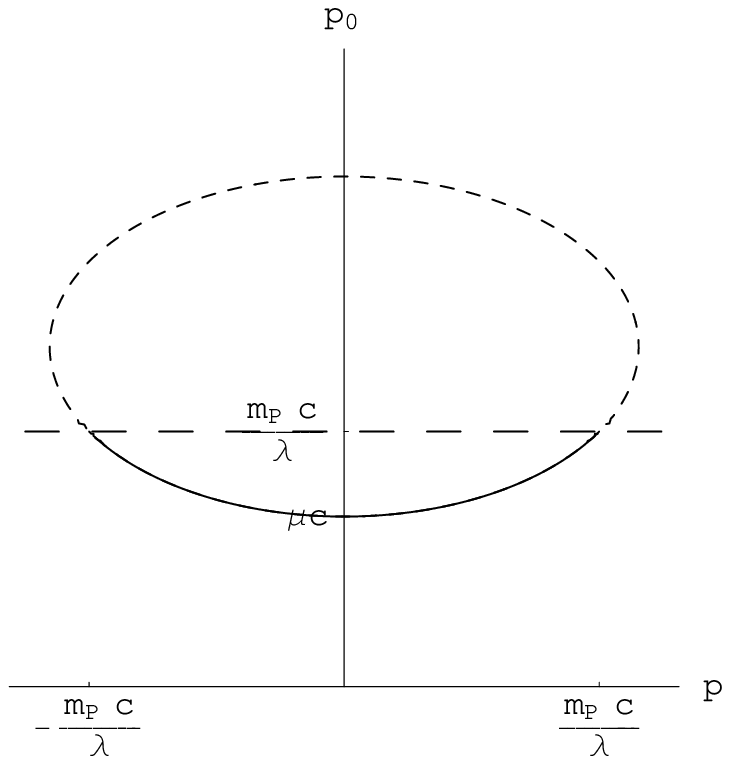}&
    \includegraphics[width=.24\textwidth]{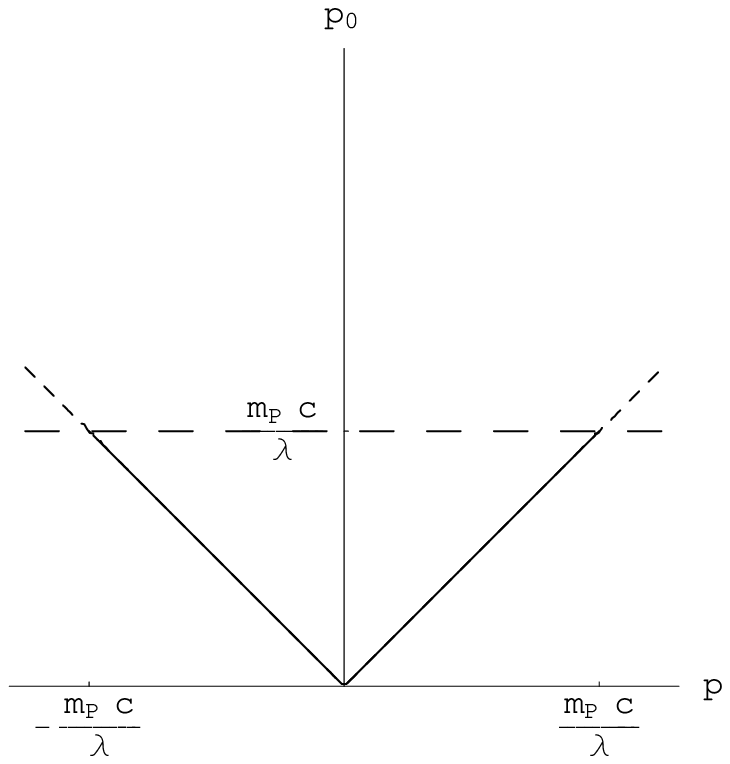}
  \end{tabular}
  \caption{Possible positive solutions of the dispersion 
    relation~\eqref{eq:4} for: (a)~$m < m_P/\lambda$, (b)~$m = m_P/\lambda$,
    (c)~$m > m_P/\lambda$, (d)~$m = 0$; $\mu = m/(1 + \lambda m/m_P)$ is the mass of
    the particle (i.e.\ the minimum of the energy).  The momentum $p$
    and energy $c p_0$ are bounded: $-m_P c/\lambda < p < m_p c/\lambda$, $\mu c \leq
    p_0 < m_P c/\lambda$}
  \label{fig:1}
\end{figure*}

Under physical condition that for the free motion the energy $c p_0$
is bounded from below we obtain four possible situations showed in the
Fig.~\ref{fig:1}.
The line $p_0 = m_P c/\lambda$ divides each curve into distinct parts: upper
and lower ones.  The momentum manifold must be an orbit of the Lorentz
group realized as in~\eqref{eq:3}.  We can easily prove that this is
possible only for the lower parts of conics showed in the
Fig.~\ref{fig:1}(a,b,d); on the upper parts of these conics the
Lorentz group cannot be globally realized.  Moreover, the upper part
of the curve showed in the Fig.~\ref{fig:1}(c), despite of the fact
that it forms an orbit of the Lorentz group, is also unphysical
because the energy of the particle reaches the maximum when its
momentum vanishes.  Furthermore, if we restricts ourselves to these
physically acceptable parts, denoted in the Fig.~\ref{fig:1} by the
solid lines, we can linearize the transformations~\eqref{eq:3} by the
following map \cite{lukierski02}
\begin{equation}
  \label{eq:6}
  k_\mu = \frac{p_\mu}{1 - \frac{\lambda}{m_P c} p_0},
\end{equation}
where the momentum components $k_\mu$ form the usual Minkowski vector.
Note, that the Jacobian determinant of this transformation vanishes
only for $p_0 = m_P c/\lambda$.  under the action of the Lorentz group
$k_\mu$'s transform linearly
\begin{subequations}
  \label{eq:7}
  \begin{align}
    k_0' &= k_0 \cosh \xi + k \sinh \xi,\\
    k' &= k \cosh \xi + k_0 \sinh \xi,
  \end{align}
\end{subequations}
and leaves the invariant $k_0^2 - k^2 = m^2 c^2$ unchanged.  Note,
that the minima of $p_0$ and $k_0$ (i.e.\ the masses) are related by
$\mu = m/(1 + \lambda m/m_P)$.

Concluding the above discussion: The nonlinear transformation
law~\eqref{eq:3} is a consequence of the choice of a nonlinear
coordinate system in the momentum space and it is not
\emph{essentially nonlinear realization of the Lorentz group}
\footnote[1]{Strictly speaking, the nonlinearity is an illusion in
  this case: As it is well known \cite{coleman69,callan69} essentially
  nonlinear group realizations are connected with the group action in
  the homogeneous coset spaces $G/H$, where $H$ is a subgroup of the
  group $G$.}, because it can be linearized by the appropriate choice
of coordinates~\eqref{eq:6}.


The question if the momentum coordinates $p_0$ and $p$ are physically
admissible is open.  In particular we do not have the notion of the
inertial frame (observer) defined operationally.  Therefore, up to
now, we cannot properly answer this question (in contrast to the
statement in e.g.\ \cite{magueijo02,judes02}).  However, if we agree
that canonical formalism can be used in such a simple kinematical
problem (i.e.\ the free motion), we can try to partially answer this
question.  According to \cite{magueijo02,judes02} let us identify the
energy with the generator of translation in time (Hamiltonian), i.e.
$H = c p_0$, while $p$ with the canonical momentum.  Therefore the
particle velocity $v$ can be calculated from one of the Hamilton
equations
\begin{equation}
  \label{eq:8}
  v = \frac{\partial H}{\partial p} = \frac{p c}{p_0 \left(1 - \frac{\lambda^2 m^2}{m_P^2}\right) 
    + \frac{\lambda^2 m^2 c}{m_P}}.
\end{equation}

Now, with help of \eqref{eq:3}, \eqref{eq:4} and \eqref{eq:8} we are
able to find the transformation law for the particle velocity under
the action of the Lorentz group
\begin{widetext}
\begin{equation}
  \label{eq:9}
  v' = \frac{c \left[\frac{v}{c} \left(\cosh \xi + \frac{v}{c} \sinh \xi\right)
      + \frac{m_P}{\lambda m} \sqrt{1 - \frac{v^2}{c^2} \left(1 
          - \frac{\lambda^2 m^2}{m_P^2}\right)}
      \left(\frac{v}{c} \cosh \xi + \sinh \xi\right)\right]}{1 - \frac{v^2}{c^2} 
    + \frac{v}{c} \left(\frac{v}{c} \cosh \xi + \sinh \xi\right) 
    + \frac{m_P}{\lambda m} \sqrt{1 - \frac{v^2}{c^2} \left(1 
        - \frac{\lambda^2 m^2}{m_P^2}\right)}  \left(\cosh \xi 
      + \frac{v}{c} \sinh \xi\right)}.
\end{equation}
\end{widetext}
The transformation law~\eqref{eq:9} goes to the standard Lorentz
transformation law for velocities when $\lambda \to 0$ or when $m = 0$ ($\mu =
0$) and $v = c$.  However, for $\lambda \neq 0$ and $m \neq 0$ it depends on the
particle mass $\mu$ (recall that $m = \mu/(1 - \lambda \mu/m_P)$).  Such a
situation is obviously in conflict with our physical space-time
intuition: Indeed, let us imagine two bodies $A$ and $B$ with mass
$m_A$ and $m_B$, respectively, moving in an inertial frame with the
same velocity $v$ (for simplicity one can assume $v = 0$).  From the
point of view of Lorentz boosted observer (i.e.\ from another inertial
frame), if $m_A \neq m_B$ and $\lambda \neq 0$, these two bodies \emph{have
  different velocities}!  This is very undesirable feature from the
physical point of view.

To understand better this issue let us look for the relationship of
$v$ and the Lorentz velocity (i.e.\ the velocity which appears in the
usual, \emph{linear}, Lorentz transformations) $v_L = c k/k_0$.  By
means of \eqref{eq:6} and \eqref{eq:8} we get that
\begin{equation}
  \label{eq:10}
  v = \frac{v_L}{1 + \frac{\lambda m}{m_P} \sqrt{1 - \frac{v_L^2}{c^2}}}.
\end{equation}
The velocity $v_L$ has the proper physical interpretation as the
velocity canonically related to $k$, i.e.\ $v_L = \partial k_0/\partial k = k/k_0$.
For the other hand it is defined as $v_L = dx/dt$, where $t$ and $x$
are usual Minkowskian time and coordinate.  However, $v_L$ cannot be
canonically related to the momentum $p$.  Notice that $v_L = k/k_0 =
p/p_0$ and it has the standard transformation law.

Concluding, we have serious problems with the velocity transformation
law and with the definition of the inertial observers within the model
presented in \cite{magueijo02}.

The next difficulty occurs when we try to formulate statistical
mechanics and thermodynamics within the framework of theory proposed
by Magueijo and Smolin \cite{magueijo02}.  Let us return to the
three-dimensional case.  It is easy to see that the invariant measure
in the momentum space is given by the following formula
\begin{widetext}
\begin{equation}
  \label{eq:11}
  d\Gamma = \frac{\left(1 - \frac{\lambda^2 m^2}{m_P^2}\right)^3 d^3p}{2 \left[1 
      - \frac{\lambda}{m_P c} \sqrt{m^2 c^2 + |\vec{p}|^2 \left(1 
          - \frac{\lambda^2 m^2}{m_P^2}\right)}\right]^3 
    \sqrt{m^2 c^2 + |\vec{p}|^2 \left(1 - \frac{\lambda^2 m^2}{m_P^2}\right)}},
\end{equation}
\end{widetext}
(this holds also for the case $m = \mu = 0$) and, since the momentum
space is bounded (see Fig.~\ref{fig:1}), the one-particle partition
function is
\begin{equation}
  \label{eq:12}
  Z_1 = V \iiint_{-\frac{m_P c}{\lambda}}^{\frac{m_P c}{\lambda}} d\Gamma\, e^{-\beta E}.
\end{equation}
But the measure~\eqref{eq:11} is singular at $|\vec{p}| \to m_P c/\lambda$ and
the energy approaches the limit $m_P c^2/\lambda$ when $|\vec{p}| \to m_P
c/\lambda$, so the integral~\eqref{eq:12} is divergent.  Consequently the
partition function does not exist in the considered case, as well as
the internal energy and the entropy.  The same holds in arbitrary
$N$-particle case.

If we take non-relativistic measure in the momentum space $d^3p$
instead of $d\Gamma$, the integral~\eqref{eq:12} becomes convergent, but on
the other hand the energy of the free gas of $N$ identical particles
(either massive or massless) is equal to \cite{judes02}
\begin{equation}
  \label{eq:13}
  E = m_P c^2 \frac{\sum_{i=1}^N \frac{E_i}{m_P c^2 - \lambda E_i}}{1 
    + \sum_{i=1}^N \frac{\lambda E_i}{m_P c^2- \lambda E_i}}
\end{equation}
and
\begin{equation}
  \label{eq:14}
  E \approx \frac{m_P c^2}{\lambda}
\end{equation}
for large $N$ \footnote{Note, that for the choice $\lambda = 1$, as it is
  assumed in \cite{magueijo02}, the mass of any large enough system is
  nearly equal to the Planck mass, i.e.\ to $2.176 \times
  10^{-8}\,\mathrm{kg}$.  On the other hand, there is no other natural
  physical choice for the parameter $\lambda$.}.  The partition function $Z$
of the $N$-particle free gas
is then of the form
\begin{equation}
  \label{eq:15}
  Z \approx e^{-\beta \frac{m_P c^2}{\lambda}} \left[8 \left(\frac{m_P c}{\lambda}\right)^3 V\right]^N
\end{equation}
and, consequently, we come at the curious conjecture that the internal
energy in the thermodynamical limit (i.e.\ for $N \to \infty$)
\begin{equation}
  \label{eq:16}
  U = -\frac{\partial\ln Z}{\partial\beta} = \frac{m_P c^2}{\lambda}
\end{equation}
does not depend on temperature!

Concluding this point of the discussion we have to state that
statistical mechanics and/or thermodynamics do not exist within the
model presented in \cite{magueijo02}.

We would like to point out that the non-additivity of energy seems to
be a common feature of doubly special relativity theories (see
\cite{judes02}).  Therefore, the question of the possiblity
of formulation of statistical mechanics or thermodynamics in such
theories is still open \footnote{In the paper \cite{kowalski01} only
  the one-particle partition function for the massless case is
  discussed within the model proposed in
  \cite{amelino01a,amelino01b,amelino01c,amelino02}.}.

We can conclude that the model proposed by Magueijo and Smolin
\cite{magueijo02} encounters some serious difficulties discussed
above.  First of all, they are interpretational problems of space-time
quantities: coordinates, velocities, etc.  Moreover, it seems to be
impossible to formulate resonably statistical mechanics and
thermodynamics within this framework.
\begin{acknowledgments}
  The Authors are grateful to J. Lukierski and J.  J{\k e}drzejewski for
  interesting discussions.  This work is supported by {\L}{\'o}d{\'z} University
  grant No.~267.
\end{acknowledgments}


\end{document}